\documentclass[twocolumn,prd,aps,showpacs,nofootinbib,nobibnotes,superscriptaddress,preprintnumbers,dvipsnames]{revtex4-2}

\usepackage{amsmath}
\usepackage{amssymb}
\makeatletter\AtBeginDocument{\let\@elt\relax}\makeatother
\usepackage[caption=false]{subfig}
\usepackage{graphicx}
\usepackage{amsfonts,amsthm}
\usepackage{xspace}
\usepackage{xcolor}
\usepackage{hyperref}

\newcommand{\Lc}{\ensuremath{\Lambda_c^+}\xspace}
\newcommand{\bendone}{\ensuremath{\theta_{C1}}\xspace}
\newcommand{\bendtwo}{\ensuremath{\theta_{C2}}\xspace}
\newcommand{\lengthone}{\ensuremath{L_{C1}}\xspace}
\newcommand{\lengthtwo}{\ensuremath{L_{C2}}\xspace}
\newcommand{\Lone}{\ensuremath{L_1}\xspace}
\newcommand{\Ltwo}{\ensuremath{L_2}\xspace}
\newcommand{\lindone}{\ensuremath{\theta_{L1}}\xspace}
\newcommand{\lindtwo}{\ensuremath{\theta_{L2}}\xspace}
\newcommand{\murad}{\ensuremath{\,\mu\text{rad}}\xspace}
\newcommand{\mrad}{\ensuremath{\,\text{mrad}}\xspace}

\newcommand{\cm}{\ensuremath{\,\text{cm}}\xspace}
\newcommand{\mm}{\ensuremath{\,\text{mm}}\xspace}
\newcommand{\mum}{\ensuremath{\,\mu\text{m}}\xspace}

\begin{document}

\title{Improved experimental layout for dipole moment measurements at the LHC}

\author{V.M. Biryukov}
\email{Valery.Biryukov@ihep.ru}
\affiliation{Institute for High Energy Physics. Protvino, 142281, Russia}
\author{J. Ruiz-Vidal}
\email{Joan.Ruiz@ific.uv.es}
\affiliation{IFIC, Universitat de Valencia - CSIC. Valencia, Spain}

\begin{abstract}

The electric and magnetic dipole moment of charm and bottom baryons can be measured for the first time by using bent crystal technology at the LHC. The experimental method, proposed in recent years, suffers from limited statistics, which dominates the uncertainty of the measurement. In this work, we present an alternative experimental layout, based on the use of crystal lenses, that improves the trapping efficiency by about a factor 15 (35) for a 2-cm (5-mm) target with respect to the nominal layout, with plain crystal faces. The efficiencies are evaluated taking into account the constraints from the LHC machine, and the technical challenges to realize this novel experimental method are discussed.

\end{abstract}

\maketitle

\section{Introduction}

Searches for permanent electric dipole moments (EDMs) of particles provide stringent constraints on theories of CP violation beyond the Standard Model. These bounds are mostly dominated by the experimental limits on the neutron and electron EDM~\cite{Pospelov:2005pr,Engel:2013lsa}. However, recent experimental results on the B-anomalies~\cite{LHCb:2019hip,LHCb:2021trn,Buttazzo:2017ixm}, CP-violation in charm~\cite{LHCb:2019hro,Bause:2020obd,Cheng:2019ggx}, or the muon $(g-2)$ \cite{Muong-2:2021ojo,Aoyama:2020ynm} hint to new physics models where the additional interactions are specific to the quark and lepton family. In these models also heavy quark EDMs can set powerful constraints~\cite{Gisbert:2019ftm}, hence the direct search for EDMs of heavy baryons are especially motivated and experimental ideas are called for. 
On the theory side, the description of charm and bottom baryon EDMs is being developed within the framework of baryon chiral perturbation theory~\cite{Unal:2020ezc,Unal:2021lhb}, while input from lattice QCD is still necessary for the interpretation of the measurement in terms of quark effective operators.

In recent years, a novel experimental method has been proposed to measure the electric and magnetic dipole moment of charm and bottom baryons\footnote{We will center the discussion on the \Lc case, while similar conclusions will obtained below for $\Xi_c^+$, and b-baryons $\Xi_b^+$ and $\Omega_b^+$.} at the LHC~\cite{Botella:2016ksl,Fomin:2017ltw}. The experimental setup, which has also been extended to measure the magnetic moment ($g-2$) of the tau lepton~\cite{Fomin:2018ybj,Fu:2019utm}, relies on the spin precession of positively charged particles, produced in a fixed target, and subjected to the electric field between the atomic planes of a bent crystal. Provided that the initial particle direction is parallel to the atomic planes, and the momentum of the particle is below some threshold, determined by the crystal bending radius, then the particle is trapped between the atomic planes, 
and it follows the curved trajectory of the bent crystal according to a probability distribution in a process called \textit{channeling}~\cite{book}. The particle has to be aligned with respect to the atomic planes within few \murad (Lindhard angle), which reduces the statistics for this measurement.

\begin{figure}[t]
	\centering
	\includegraphics[width=0.95\linewidth]{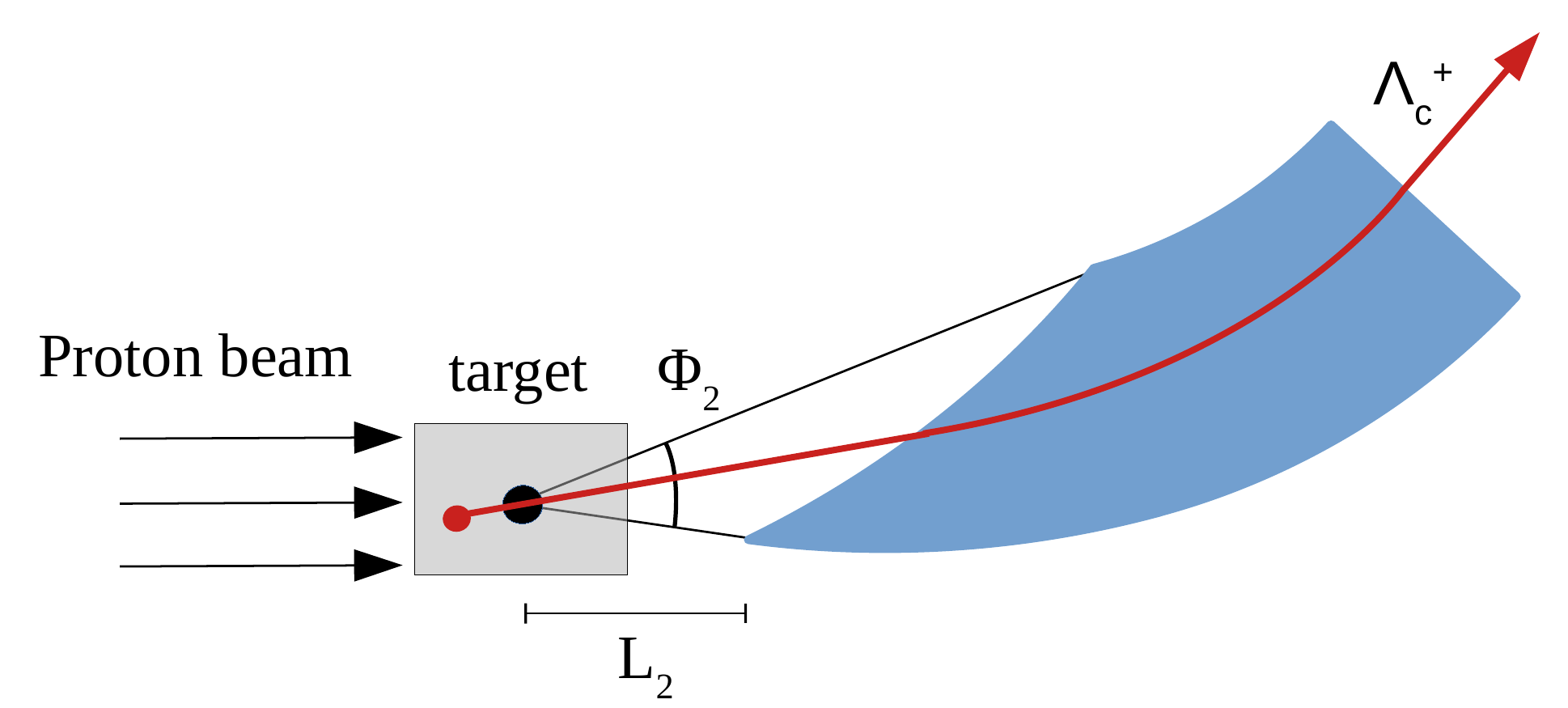}
	\caption{Sketch of the single-lens scheme. The proton beamlet arrives directly to the target, where the \Lc are produced (red dot) and a small fraction ($\sim 10^{-3}$) of them are subsequently channeled in the crystal lens.}
	\label{fig:generaldiagramsinglelens}
\end{figure}

\begin{figure*}[t]
	\centering
	\includegraphics[width=0.7\textwidth]{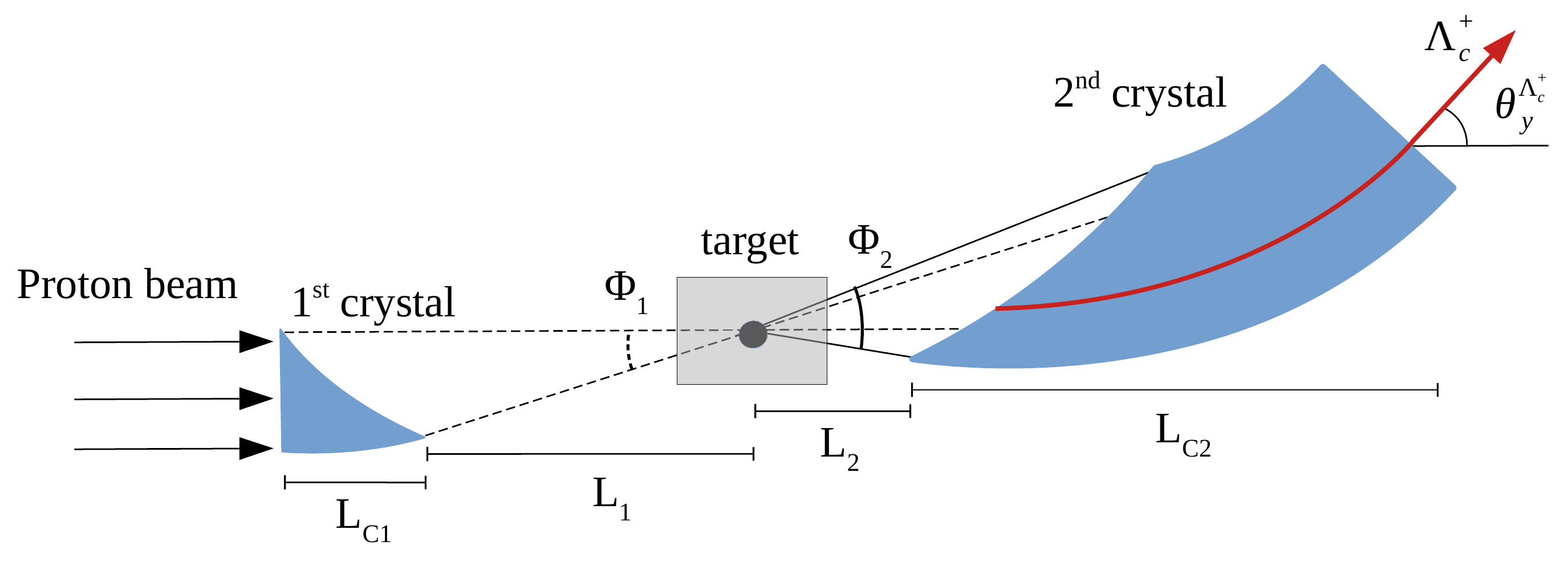}
	\caption{The protons are focused onto the target by means of a crystal lens of length \lengthone, at a distance \Lone from the center of the target. The interaction products, such as $\Lc$ particles (red curve), may be channeled in a second crystal lens if their direction is parallel to the atomic planes at the crystal entry point. To capture particles produced at different angles, the lens aperture of the second crystal, $\Phi_2$, must be at least that of the first crystal plus the maximum opening angle of the \Lc, $\Phi_2 \approx \Phi_1 + 2 \Phi_{\Lc}$. 	The maximum deflection angle $\bendtwo$ is experienced by the \Lc particles that are produced with an aperture $-\Phi_{\Lc}$, below the horizontal. After being fully deflected by an angle $+\bendtwo$, these exit the crystal lens at an angle $\theta_y^{\Lc} = \bendtwo-\Phi_{\Lc} \approx  \bendtwo - (\Phi_2 - \Phi_1)/2$. 
	}
	\label{fig:generalDiagram}
\end{figure*}

While the Lindhard angle depends on the crystal material and the particle energy, and cannot be increased to trap particles in a wider range of directions, the overall geometry of the setup can be composed in such a way that the atomic planes are laid out in different directions, matching the initial angle of more of the produced particles~\cite{Biryukov:2021gsd,Biryukov:2021phs}. 
This idea can be realized by replacing the plain crystal of the nominal setup by a single crystal lens, as shown in Fig.~\ref{fig:generaldiagramsinglelens}. However, to reach a radical increase on the trapping efficiency, the initial proton beam must be focused onto the target by means of a first lens, as shown in Fig.~\ref{fig:generalDiagram}.

A similar scheme, with a crystal deflector and two lenses with coinciding focal points was already realized at the IHEP experiment with 70-GeV protons~\cite{Baranov:1995sx}. The purpose of that experiment was to show that a whole beamline can be realized with crystal elements where magnet dipoles and quadrupoles are replaced with crystal deflectors and lenses. This crystal "beamline" channeled $10^7$ proton/s. The second lens trapped a beam with divergence 100 times greater than the Lindhard angle with an efficiency of 15\%, \textit{i.e.} much higher than would be the efficiency of a plain crystal in this divergent beam. First tests of crystal lenses at IHEP~\cite{Denison:1991vf} have shown that the focused beam followed the theory expectations, with the beam size at focus measured to be the Lindhard angle times the focal distance, as it should be.

The use of crystal lenses for spin precession experiments represents a new application of the bent-crystal channeling technique, which was tested at many accelerators \cite{dubna,sps,fnal-tev,ihep-plb,rhic-pr,kek} and routinely exploited since the 80's at IHEP for beam delivery to particle physics experiments \cite{ufn} at intensities up to $10^{12}$ proton/s. At this intensity, the channeling efficiency reached 85\% \cite{ihep-nim}. Six locations on IHEP 70-GeV ring are equipped by crystal extraction systems, serving for applications rather than for research \cite{ihep-nim}. At much higher proton energies, simulations predicted high efficiency for LHC beam extraction \cite{prl} and collimation \cite{nim-lhc}, while recent experiments demonstrated that bent crystals can be very efficient at LHC energies \cite{lhc}.

The paper is organized as follows. In Section~\ref{sec:scheme}, the proposed experimental solutions are described in detail, pointing out the differences with the originally proposed plain-crystal scheme. The compatibility of this setup with the LHC beam is addressed in Section~\ref{sec:parameters}, where realistic values for the setup parameters are obtained. These are used subsequently in Section~\ref{sec:efficiency} to evaluate the trapping efficiency, always in comparison to the plain-crystal scheme. The technical limitations for the single- and double-lens scheme are briefly discussed in Section~\ref{sec:technology} before summarizing our results in Section~\ref{sec:conclusion}.

\section{Experimental layout} \label{sec:scheme}

In the proposed experiment, based on the plain-crystal scheme~\cite{Botella:2016ksl,Bagli:2017foe,Aiola:2020yam,Fomin:2017ltw,Fomin:2019wuw,Baryshevsky:2018hvh}, part of the protons from the LHC beam halo are deflected at an angle of about $100\murad$ with a crystal \textit{kicker} positioned about 100 m upstream of the target, creating a secondary beamlet within the LHC beam pipe. Accurate simulations of the beam optics project these protons to be about 4 \mm away from the main beam when they arrive to the fixed-target setup~\cite{Mirarchi:2019vqi}, positioned either before the LHCb detector or at the IR3 interaction region, in a future dedicated experiment. At this point, the 7 TeV protons interact with a tungsten target of $\approx 2\cm$ with a center-of-mass energy of 115  GeV, producing copious amounts of charm hadrons in inelastic collisions. The electric and magnetic moment of the \Lc baryons can be measured using the events in which these particles are channeled in a bent crystal attached to the target. A net electromagnetic field deflects their trajectory and induces the precession of the \Lc magnetic moment around the magnetic field, as seen in the reference frame of the particle. 
Any change on the polarization in the perpendicular direction to the main precession, \textit{i.e} around electric field, signals the presence of a nonzero EDM, at least at the sensitivity level of the planned experiments~\cite{Baryshevsky:2018hvh}. At the crystal exit, these \Lc particles reach an angle with respect to the proton beam sufficient for its decay products to enter in the detector acceptance, whereas other products of the proton-target interactions are produced in the forward direction, and are contained within the LHC beam pipe. Finally, the \Lc polarization is reconstructed through the angular distribution of \textit{e.g.} $\Lc \to p K^- \pi^+$ decay~\cite{Marangotto:2020ead}. The projected uncertainty on this method is limited by the available statistics, which is determined by the small number of protons that can be deflected from the LHC beam halo, and the low trapping efficiency of the crystal for the \Lc particles produced at the target. The interesting \Lc candidates, with $E\gtrsim700\text{GeV}$, are produced approximately in a cone of $\pm 1.5 \mrad$, while only those that are produced aligned to crystal planes, within an angle of $\pm 7\murad$ (Lindhard angle), are trapped between the crystal atomic planes. \footnote{We will take the Lindhard angle $\lindtwo = 7 \murad$ in both schemes, corresponding to 700-GeV particles in a Silicon crystal oriented in the (110) direction. Since the efficiency is always computed relative to the plain-crystal scheme, this assumption cancels out to a large extent.}

To increase this efficiency one could consider to replace the plain crystal by a crystal lens, whose atomic planes point towards the target. Naively thinking, this would allow to capture all \Lc produced at the target. However, in reality the target volume is much larger than the focal volume, \textit{i.e.} the region in which the produced \Lc can be trapped into the crystal.  
Thus, in qualitative terms, while the plain crystal traps particles produced across the whole target volume but restricted to a small range of directions, the crystal lens traps the particles from all directions, but restricted to a small volume. To perform a quantitative comparison of the two schemes, we shall derive the trapping condition for the crystal lens anew.
The basic requirement is that the \Lc direction must be within the Lindhard angle of the atomic planes ($\theta_{L2}$ in Fig. \ref{fig:crystalParameters}) when it reaches the crystal entry face. The direction of the atomic planes varies across the crystal entry face, but all of them point towards the focal point, where the projection of the Lindhard angle defines a \textit{focal window}. Thus, the trapping condition can be redefined in a more practical way: the (extended) \Lc trajectory must cross the focal window for it to be trapped in the crystal. The height of the focal window can be determined with the help of Fig.~\ref{fig:crystalParameters} and reads
\begin{equation}
w_{F2} = 2\lindtwo (\Ltwo + l_2) ~,
\label{eq:window}
\end{equation} 
where \Ltwo is the target-crystal distance and $l_2$ depends on the entrance point, as defined in Fig. \ref{fig:crystalParameters}. This height $w_{F2} \approx 0.4-0.7 \mum$ has to be compared to the beam size $\sigma_{\rm beam} \approx 50 \mum$ on the $y$ direction. Thus, in the single-lens scheme, only few of the protons are approximately aligned towards the focal window, reducing the probability of the particles produced at the target to cross it. However, even if this probability is small, so is the trapping efficiency in the nominal plain-crystal scheme, and the comparison is not straightforward. In fact, detailed evaluations of this efficiency, described in Section \ref{sec:efficiency}, show an improvement of around 40\% with respect to the plain-crystal scheme.

By reducing the beam size at the focal window this gain factor could be further increased. This can be achieved by introducing a first crystal lens prior to the target that focuses the proton trajectories towards the same focal point. In this way, all of the protons cross the focal window of the second crystal, and the probability of the produced \Lc to be trapped increases drastically. We shall evaluate the trapping efficiency of the double-lens scheme in Section \ref{sec:efficiency}, after realistic values for the setup parameters are obtained in section \ref{sec:parameters}.

\section{Setup parameters and detector constraints} \label{sec:parameters}

\begin{figure}[t]
	\centering
	\includegraphics[width=0.99\columnwidth]{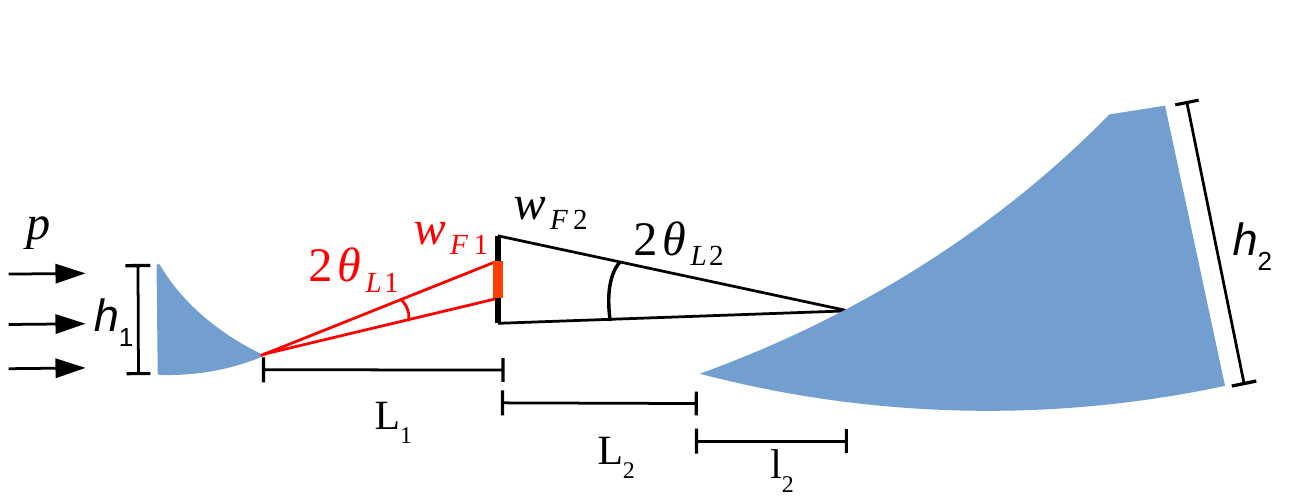}
	\caption{The extended trajectory of the particles must cross the focal window $w_{F2}$ (vertical black line) for these to be trapped in the second crystal. To maximize this probability, the focal window $w_{F1}$ (thick red line), analogously defined for the first crystal, must be embedded within $w_{F2}$.}
	\label{fig:crystalParameters}
\end{figure}

To maximize the trapping efficiency, the focal window of the first crystal lens should be embedded within that of the second lens, \textit{i.e.} $w_{F1} \lesssim w_{F2}$ (see Fig. \ref{fig:crystalParameters}). Following Eq.~\eqref{eq:window}, this condition imposes the relation between the focal lengths, $\Lone \lesssim (\lindtwo / \lindone) \Ltwo$. To avoid efficiency losses due to the exponential decay of the \Lc after its production, the second crystal should be as close as possible to the target. Taking $\Ltwo \approx 3 \cm$, then the first crystal should be positioned at $\Lone \lesssim 8 \cm$, where we used the ratio $(\lindtwo / \lindone)\approx 2.7$, for 7 TeV (1 TeV) protons (\Lc). This excludes the possibility of converting the crystal \textit{kicker} into the first crystal lens, as it will be positioned $\mathcal{O}(100 \text{ m})$ before the target.

\begin{figure}
	\centering
	\includegraphics[width=0.85\linewidth]{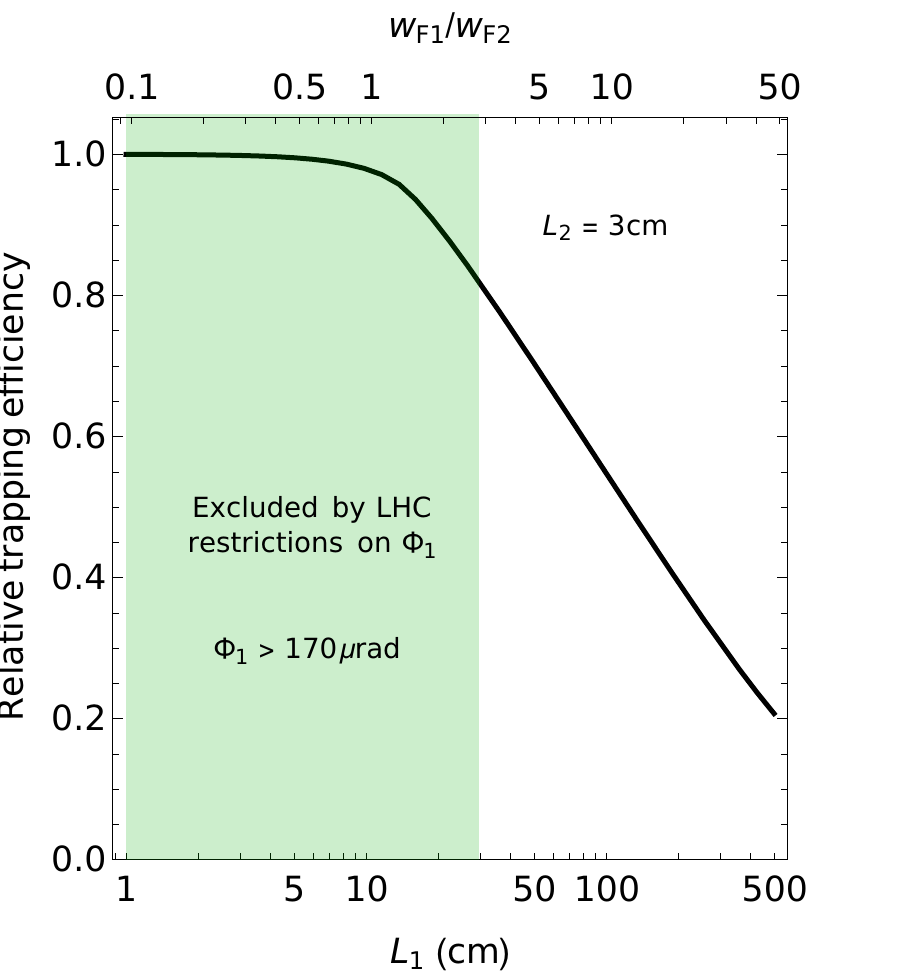}
	\caption{Trapping efficiency of the double-lens scheme as a function of the distance $L_1$ and, on the upper scale, of the ratio of focal windows $w_{F1}/w_{F2}$. The two scales are related via Eq. \eqref{eq:window}. When the focal window of lens 1 is embedded in that of lens 2, \textit{i.e.} when $w_{F1}/w_{F2}<1$, there is a plateau on the trapping efficiency. Increasing this ratio up to $w_{F1}/w_{F2}=2.8$, to make the setup compatible with the LHC beam pipe, reduces the efficiency to 82\% of the initial value.}
	\label{fig:effL1}
\end{figure}

Regarding the focusing angle of the two lenses, $\Phi_1$ and $\Phi_2$ in Fig.~\ref{fig:generalDiagram}, there are critical requirements on the beamlet position and size for it to be compatible with the LHC beam pipe. 
The divergent proton beamlet after the first crystal lens should be stopped by an absorber while it is still contained within the beam pipe. Imposing a maximum width of $1\cm$ at the absorber position, around 60~m downstream of the setup~\cite{Barschel:2020drr}, then the divergence of the first crystal is restricted to be $\Phi_1 \lesssim (1\cm/60\,{\rm m}) \approx 170 \murad$.
This, however, is incompatible with the previous considerations about the focal distance \Lone: 
with such small focusing angle, the needed distance to focus a beam of width $\sigma_{\rm beam}$ is $\Lone = \sigma_{\rm beam} /\Phi_1 = 30 \cm$.
In turn, increasing $L_1$ from 8 to 30 \cm would spread the proton directions due to the Lindhard angle, making $w_{F1}>w_{F2}$. However, in contrast to the qualitative arguments given above, evaluating the efficiency for $L_1=30\cm$ we see that it induces only a mild effect on the trapping efficiency, whose dependence on $L_1$ is far from proportional as shown in Fig. \ref{fig:effL1}. This efficiency was evaluated following the procedure described in Section~\ref{sec:efficiency}. The resulting efficiency loss, of 18\%, could be avoided either by reducing the distance to the beam absorber; reducing the transverse size of the beamlet before the first crystal lens; increasing the Lindhard angle of the second crystal lens \textit{e.g.} using a Germanium crystal; or a combination of some of these possibilities, reaching an overall factor of 3. However, to be conservative in the evaluation of the trapping efficiency we will not consider any of these improvements, keeping the absorber position unchanged.

To capture all the \Lc particles crossing the focal window with any angle $\theta$ with respect to the impinging proton, the focusing angle of the second crystal lens should cover all possible proton directions plus some extra room for the aperture angle of the \Lc, \textit{i.e.} $\Phi_2 = \Phi_1 + 2 \Phi_{\Lc}$, where $\Phi_{\Lc} \approx 1.5 \mrad $ is the maximum aperture of the $\Lc$ particles with respect to the incoming proton direction (see Fig. \ref{fig:generalDiagram}). Note that the distribution of proton angles within $\Phi_1$ is approximately uniform.
Thus the focusing angle of the second crystal should be $\Phi_2 \approx 3.2 \mrad$, although it can be increased arbitrarily without affecting the trapping efficiency. 

Regarding the specifications for the two crystal lenses, the optimal crystal parameters for the plain-crystal scheme have been updated recently in~\cite{Aiola:2020yam} using state-of-the-art simulations for crystal channeling~\cite{Sytov:2014jfa,Sytov:2019gad}. An optimal crystal with $\lengthtwo=10\cm$ and $\bendtwo=16\mrad$ was found to maximize the sensitivity to the electric and magnetic dipole moments.
The minimum crystal bending angle is determined by the acceptance of the detector, which imposes a minimum angle for the \Lc away from the beam pipe, $\theta_y^{\Lc} = \bendtwo$ . However, bringing this same bending angle to the double-lens scheme slightly decreases the angle of the outgoing \Lc with respect to the horizontal, being $\theta_y^{\Lc} = \bendtwo - (\Phi_2 - \Phi_1)/2$ (see Fig. \ref{fig:generalDiagram}).
This deviation may be corrected by increasing the bending angle by an extra 1-2 \mrad, which induces a small loss on the channeling efficiency. However, this effect is roughly compensated by the reduction in dechanneling losses of the crystal lens, which also captures \Lc at positive angles, which \textit{see} a smaller crystal bending. Thus, in the comparison with the plain-crystal scheme, the same channeling efficiency is assumed for the particles already trapped within the Lindhard angle.

\begin{table}[t] \label{tab:parameters}
	\caption{Setup parameters used for the computation of the trapping efficiency with the double-lens scheme. The parameter values are either determined by the physical limitations of the LHC beam, or taken as the optimal parameters from previous studies (see text).}
\begin{tabular}{cccc}
	\hline
	Parameter & Value & Parameter & Value \\
	\hline
	\Lone & 30 \cm         &  \Ltwo & 3 \cm \\
	$\Phi_1$  & 170 \murad &  $\Phi_2$  & 3.2 \mrad \\
	\lengthone & -    &  \lengthtwo & 10 \cm \\
	\bendone & 170 \murad     &  \bendtwo & 18 \mrad \\
	\hline
\end{tabular}
\end{table}

\section{Realistic trapping efficiency} \label{sec:efficiency}

\subsection{Double-lens scheme}

\begin{figure}
	\centering
	\subfloat[]{\includegraphics[width=0.99\columnwidth]{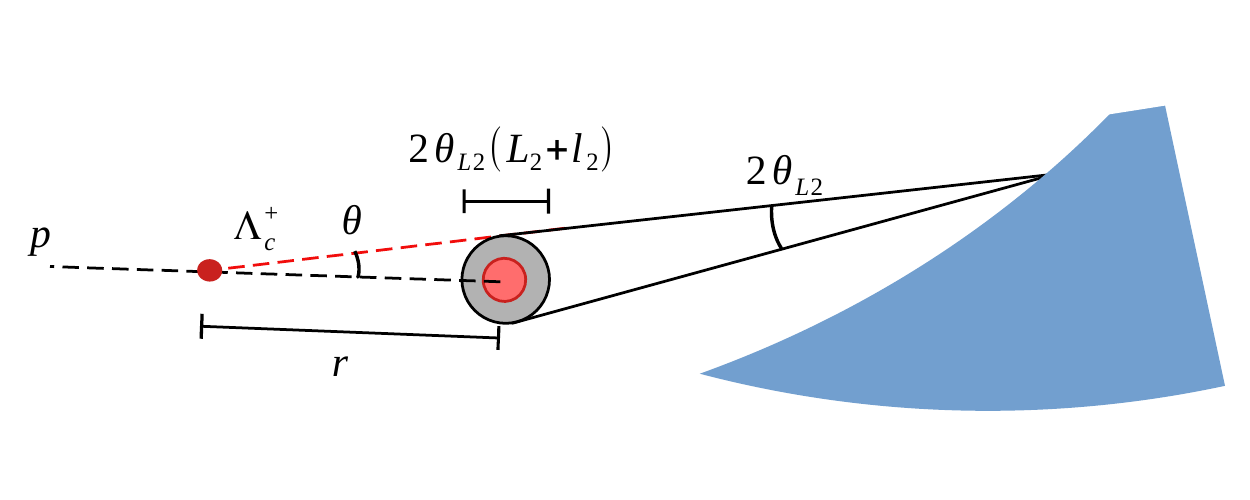}} \\
	\subfloat[]{\includegraphics[width=0.99\columnwidth]{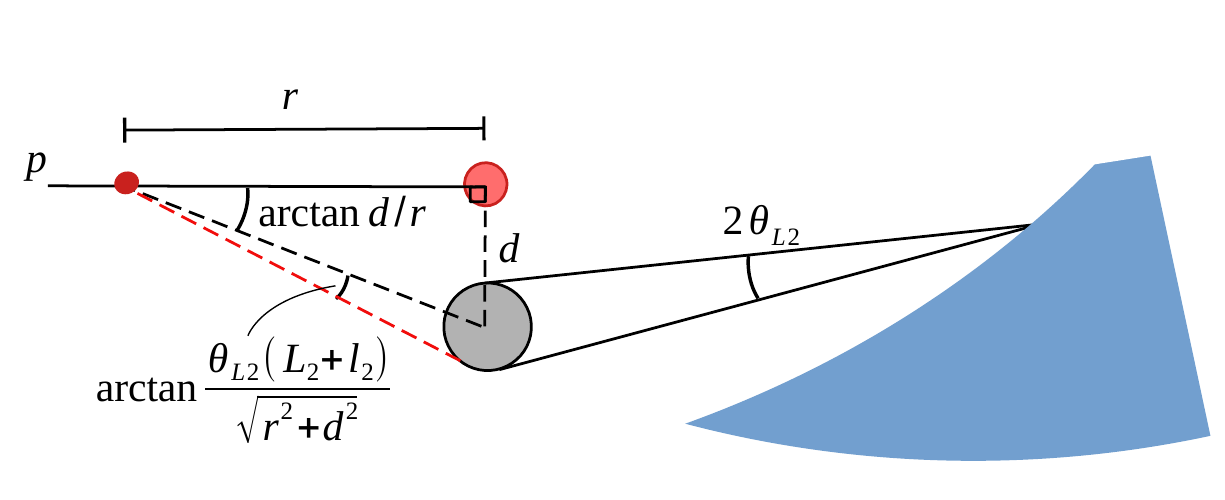}}
	\caption{If we imagine the crystal face to be equidistant to the focal point, \textit{i.e} $l_2 = 0$, then the Lindhard angle projected from every entry point of the crystal defines a circumference (in grey). The extended trajectory of the \Lc particles must overlap this circumference (cylinder in 3D) to be trapped in the crystal. For coinciding focal points (a), the trapping condition depends simultaneously on the \Lc aperture angle, $\theta$, and the distance between the production and the focal point of the first crystal lens, $r$ (both projected onto the bending plane). To define the trapping condition for vertically displaced focal points (b), we assumed the impinging proton to be horizontal, as deviations of the proton trajectory for positive and negative angles approximatley cancel out. This allows to take a right angle and extract the trapping condition with only one additional parameter: the distance $d$ between the two foci. }
	\label{fig:channelingconditions}
\end{figure}

The concept of focal window was useful to compare the different schemes and determine the setup parameters with quick estimations. However, to estimate the trapping efficiency reliably, also in the case of misaligned crystals, it is convenient to use a trapping condition based not on the focal window but on its corresponding solid of revolution. This is in fact free of small-angle approximations and allows to compute the trapping efficiency when the focal points of the two crystals are in arbitrary different positions. Following the geometrical arguments in Fig. \ref{fig:channelingconditions}a, the trapping condition can be expressed as 
$$
|r \tan(\theta)| \leq \lindtwo ( \Ltwo + l_2)~,
$$
where $\theta$ is the \Lc aperture angle with respect to the impinging proton, and $r$ is the signed distance between its production and the focal point, with positive values to the right of the focal point, and viceversa.
Taking a small $\theta$ angle approximation and a conservative value of $l_2=0$, we can compare this expression, $|\theta| \leq \lindtwo (\Ltwo/|r|) $, to the trapping condition in the plain-crystal scheme, $\theta \leq \lindtwo$, which is independent of the production point. The gain factor of the double-lens scheme is therefore $\Ltwo/|r|$, which is always greater than one and tends to infinity when $|r|\to0$. This is perfect agreement with the geometrical view presented above by which all \Lc produced at the focal point are in acceptance. Taking the average value $\langle |r| \rangle = 0.5 \cm$, the factor $\Ltwo/\langle |r| \rangle \approx 6$ already provides an order-of-magnitude estimation of the gain with respect to the plain-crystal scheme. This is, nevertheless, only a lower limit to the real gain, which grows rapidly for small $r$. The final efficiency can be evaluated as the portion of phase space fulfilling the trapping condition. To evaluate it, we shall integrate over $r$ and $\theta$ in the region
$$
R = \lbrace(r,\theta)~ : ~ (|r| \leq 1 \cm, ~|\theta| \leq \lindtwo (\Ltwo+l_2)/|r|) \rbrace~.
$$
Analogously, the available phase space in the plain-crystal scheme is determined by the phase space region 
$$
P = \lbrace(r,\theta)~ : ~ (|r| \leq 1 \cm, ~|\theta| \leq \lindtwo ) \rbrace~.
$$
To account for the angular distribution of the \Lc particles around the proton direction, we will integrate over a probability density function $G(\theta;~\mu, \sigma)$ defining a normal distribution in $\theta$ with $\mu = 0$ and $\sigma=1.5 \mrad$. This also makes sure that for $r\to0$ the efficiency stays finite. The number of events in acceptance is therefore proportional to 
\begin{equation} \label{eq:ntrappedevents}
 N_R = \int_R G(\theta;~0, ~1.5~\mrad) ~dr ~d\theta~.
\end{equation}
Note that we could include the attenuation of the proton beam within the target with an exponential density function in $r$. However, for 7 TeV protons within a tungsten target of 2 cm, the distribution in $r$ is nearly uniform, and we will leave this correction out for the sake of clarity.

\begin{figure}
	\centering
	\includegraphics[width=0.90\linewidth]{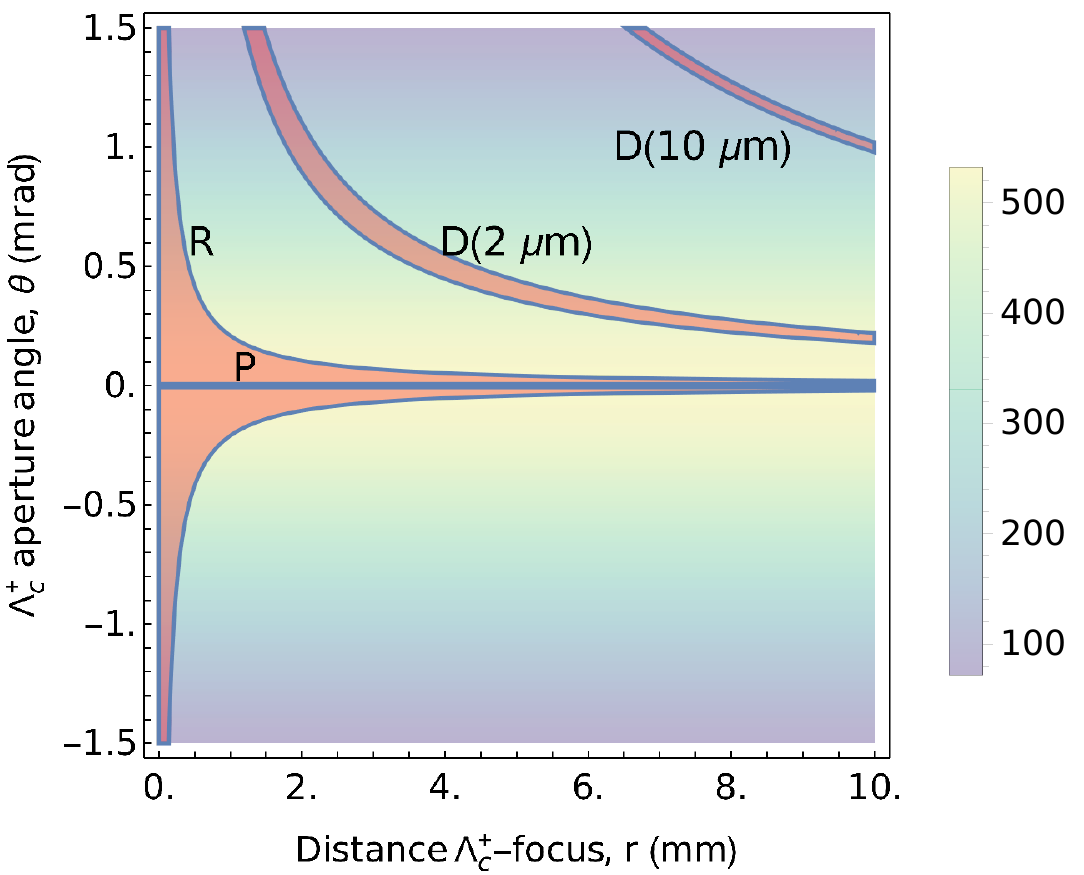}
	\caption{Regions of the phase space of the variables $(r,\theta)$ where the trapping condition is met for the plain-crystal scheme (P), the double-lens scheme (R), and the double-lens scheme with a vertical displacement between the two focal points of $d$ ($D(d)$). The background color gradient shows the probability distribution of the \Lc aperture angle, $\theta$, in Eq~\eqref{eq:ntrappedevents}. Even though the phase space for the plain-crystal scheme is smaller than the one for vertically displaced focal points, it is centered at the peak of the $\theta$ distribution, partially compensating this difference. This plot is antisymmetric for negative values of $r$.}
	\label{fig:phasespace}
\end{figure}

Since the focal windows $w_{F1}$ and $w_{F2}$ are so small, we should consider also the possible misalignment of the two crystal lenses. From Fig. \ref{fig:crystalParameters}, one can see that a displacement of the two focal points in the horizontal direction would not have a great effect on the trapping effiency, as the impinging protons and produced particles travel close to the horizontal direction. A displacement in the vertical direction is however much more critical, and it may have a large effect on the trapping efficiency. With the help of Fig. \ref{fig:channelingconditions}b we can define the trapping condition for two focal points misaligned in the vertical direction by a distance $d$,\footnote{The trapping condition for the plain-crystal scheme, $|\theta|<\theta_{L2}$, can be recovered when the focal length tends to infinity and $r\approx \Ltwo \gg d$.}
\begin{equation}\label{eq:conditionDisplaced}
|\theta - \arctan\frac{d}{r} |\leq 
 \arctan \frac{\theta_{L2}\Ltwo}{\sqrt{d^2 + r^2}} ~,
\end{equation}
where $r$ is in this case the signed distance between the proton interaction and the focal point of the first crystal lens. Note that the previous definition of $r$ is just a limiting case for coinciding focal points, as shown in Fig.~\ref{fig:channelingconditions}. The corresponding integration region will be noted by $D(d)$, which in this case depends on the vertical distance of the two foci, $d$. This region of phase space is shown in Fig.~\ref{fig:phasespace}, together with that of a perfectly aligned setup, and that of the plain-crystal scheme. As anticipated, the efficiency gain of the double-lens scheme with respect to the plain-crystal scheme is much larger than a factor 6, as seen by the difference in area between the $R$ and $P$ regions. The case for misaligned crystals suffers a rapid efficiency loss for small vertical displacements.

Before carrying out the integration and obtaining the final efficiency, there is one more effect needed to accurately estimate the trapping efficiency of the double-lens scheme. So far we considered all the proton trajectories to intersect exactly at the focal point, but due to the margin of the Lindhard angle at the crystal exit, these directions are diluted within the focal window of the first crystal, $w_{F1}$. 
To account for this smearing without Monte Carlo simulations, we evaluate the trapping efficiency for small vertical displacements $\delta$ of the proton trajectories within focal window $w_{F1}$ and take their average through the integral
\begin{equation}\label{eq:Nwindow}
N(d) = \frac{1}{w_{F1}}  \int_{-w_{F1}/2}^{w_{F1}/2} N_{D(d + \delta)}  ~\text d \delta~~,
\end{equation}
where $N_{D(d)}$ is defined as in Eq. \eqref{eq:ntrappedevents}, and $d$ is the vertical displacement between the central points of the two crystal foci. Then, using our analytical method, the number of events in acceptance is the result of integrating sequentially over the phase space that meets the trapping condition, in Eq. \eqref{eq:ntrappedevents}, and over the small displacements within the focal window of the first crystal, in Eq. \eqref{eq:Nwindow}.

\begin{figure}
	\centering
	\includegraphics[width=0.80\linewidth]{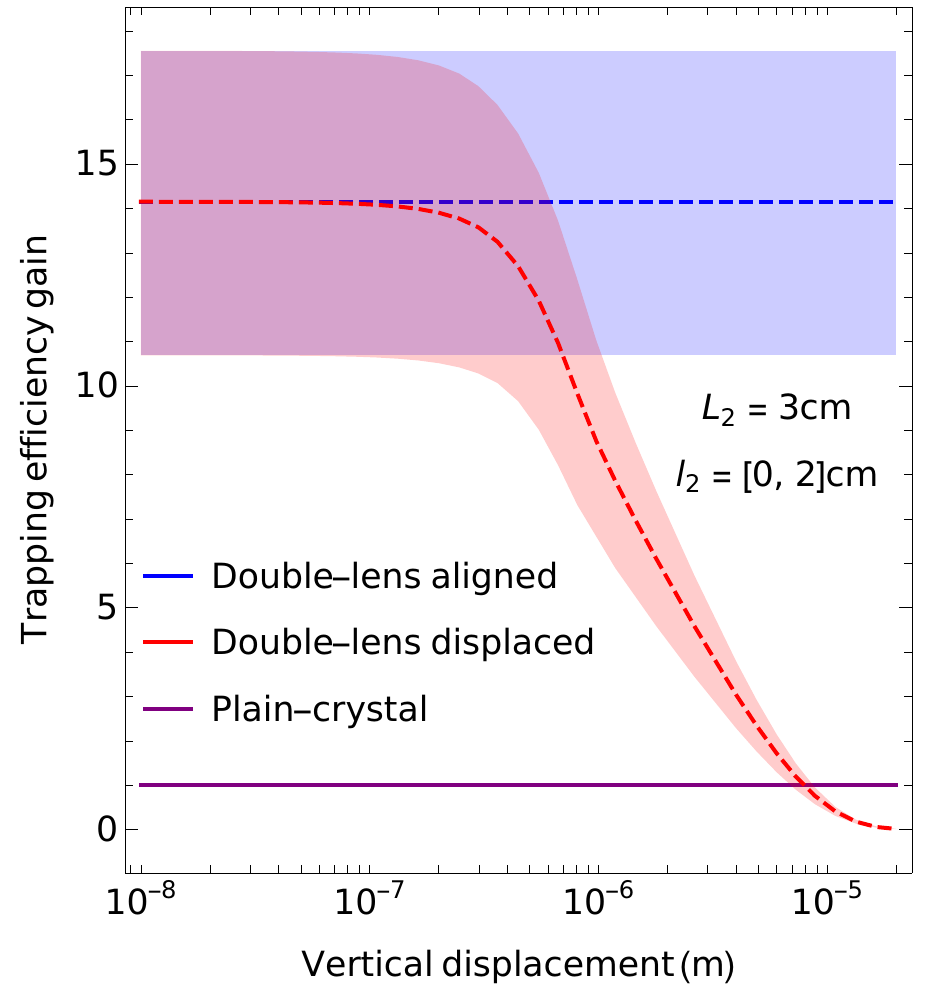}
	\caption{Trapping efficiency with respect to the plain-crystal scheme, as a function of the vertical displacement of the two focal points, with a 2-cm target. The shaded bands represent the variation of the trapping efficiency across the crystal entry face, due to the small change on the distance to the focal point, noted as $l_2$ in Fig.\ref{fig:crystalParameters}. The maximum extra distance is $l_2^{\text{max}} = \lengthtwo \Phi_2 / \bendtwo$, and the average efficiency (dashed lines) is met in the middle of the crystal entry face, with $l_2^{\rm max} /2$. 
	}
	\label{fig:trappingefficiency}
\end{figure}

Finally, the trapping efficiency gain of the double-lens scheme with respect to the plain-crystal scheme is shown in Fig. \ref{fig:trappingefficiency} as a function of the vertical displacement $d$. We used the setup parameters summarized in Table~\ref{tab:parameters} and a target thickness of 2\cm. The maximum gain reaches about a factor 15, but decreases rapidly for small vertical displacements of the two focal points, falling below the plain-crystal scheme for $d \gtrsim 8 \mum$. In principle, this accuracy is achievable with current technology and the proposed setup would not represent an efficiency loss in any case, as discussed in Section~\ref{sec:technology}.

Provided that the focal windows are made to overlap by adjusting the crystal position itself, a limited angular precision on the second crystal orientation is of no concern. It is only important that $\Phi_2$ is ample enough to contain all the particles produced at different angles, as discussed in Fig. \ref{fig:generalDiagram}. To ensure this, $\Phi_2$ may be increased slightly by some number of sigmas of this angular precision.
Also, to consider the possible inefficiencies of the first lens, the gain of the double-lens scheme, in Fig.~\ref{fig:trappingefficiency}, should be re-scaled by the channeling efficiency of protons in the first lens.
The proton bending efficiency measured at SPS was 77-83\% for bendings of $0.05-0.2\mrad$ for beams of low divergence \cite{sps-rep}. 
In our case, the beam divergence is equal to the Lindhard angle, as induced by the beam-halo crystal deflector, while the proton bending ranges from 0 to 0.17\mrad.
Therefore we expect that the first-lens efficiency can be about 60-80\%. Other small differences with the plain-crystal scheme such as dechanneling losses or the increased bending angle approximately cancel out, as explained in the previous section. 
Note also that the maximum gain factor, for perfect alignment, increases from 15 to 17 if the specific setup at the LHC allows for a beam divergence $\Phi_1 \approx 300\murad$, as shown in Fig.~\ref{fig:effL1}.

\begin{figure}
	\centering
	\includegraphics[width=0.8\linewidth]{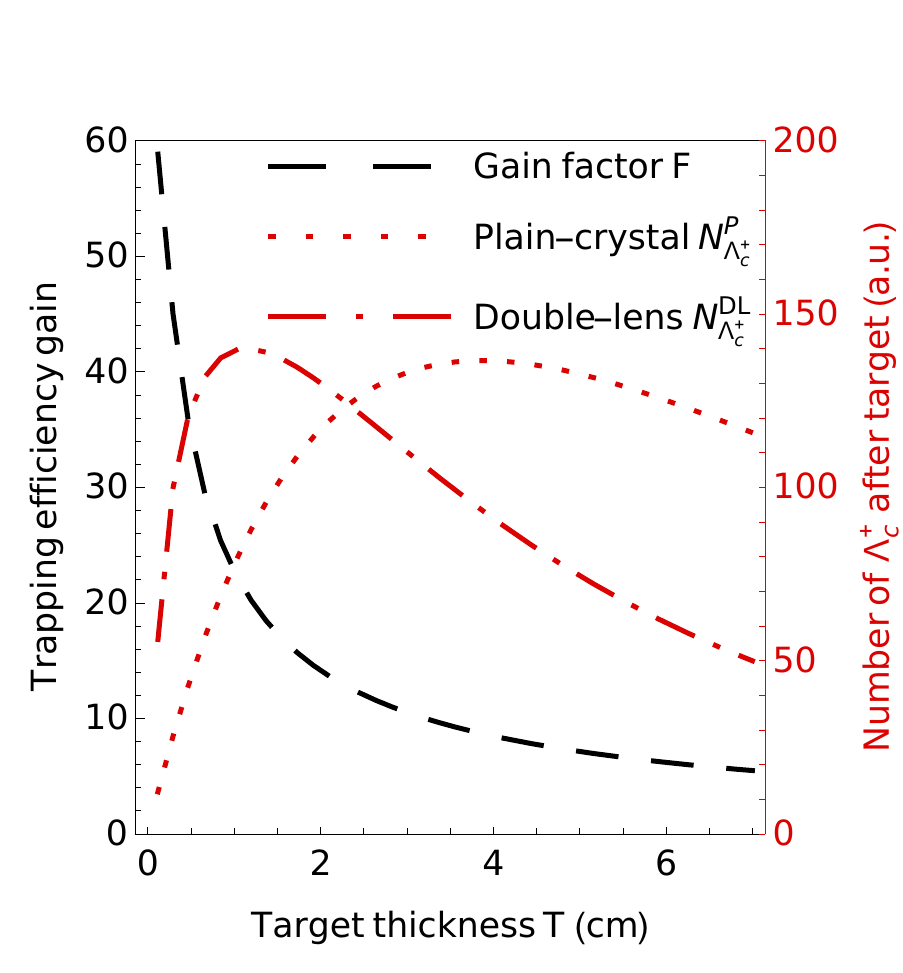}
	\caption{Showing, as a function of the target thickness, the trapping efficiency gain factor F (dashed black); the number of \Lc after the target, also valid for the plain-crystal scheme (dotted red); and the number of produced-survived-and-trapped \Lc in the double-lens scheme (dot-dashed red). The optimal target thickness for the double-lens (plain-crystal) scheme is at 0.5\cm (2.0 cm), corresponding to a gain factor $F \approx 35$.}
	\label{fig:opttarget}
\end{figure}

Moreover, any reduction on the target thickness would affect the efficiency of the double-lens scheme much less than in the plain-crystal scheme, since most of the trapped particles are produced around the center of the target, at $|r|\approx0$. For instance, if the target thickness was reduced from 2 to 0.5\cm along the beam direction, the efficiency improvement of the proposed layout would increase from a factor 15 to 35. A detailed optimization of the target thickness $T$ is presented in Fig.~\ref{fig:opttarget}. The number of \Lc after the target was evaluated as a function of $T$ in Fig.~10 of Ref.~\cite{Aiola:2020yam} accounting for the proton flux attenuation, and the absorption and decay of \Lc within the target. Reproducing this dependency for the plain-crystal scheme $N_{\Lc}^P (T)$ (dotted line), and multiplying it by the gain factor of the double-lens scheme $F(T)$ (dashed line), we obtain the variation with $T$ of the number of produced-survived-and-trapped \Lc in the double-lens scheme as $N_{\Lc}^{DL} (T) \propto F(T)~N_{\Lc}^P (T)$ (dash-dotted line). 
The maximum efficiency is found at $T=1\cm$. However, to reduce the background from proton-target interactions, we take the shortest target within a $10\%$-difference in efficiency from the optimal point, yielding $T\approx0.5\cm$.

In summary, in this section we have found that the optimal target thickness for the double-lens scheme is ${\approx0.5\cm}$. With the same target thickness, there is a gain of a factor 35 in trapping efficiency with respect to the plain-crystal scheme. Accounting for the first-lens efficiency ($\sim 80\%$) and the decay of the \Lc in the additional target-crystal separation of 2\cm in $\Ltwo$ ($\sim 70\%$), the final gain on number of events would be of a factor $\sim 20$. For bottom baryons or $\Xi_c^+$, the gain would be slightly better due to their longer lifetimes. However, to benefit from the increased efficiency of the double-lens scheme, additional technical challenges must be addressed in comparison to the plain-crystal scheme. These are briefly described in Section \ref{sec:technology}.

\subsection{Single-lens scheme}

Evaluating the efficiency of the single-lens scheme is straightforward with the tools developed for the double-lens scheme. In this case, the protons are distributed uniformly in the vertical direction, resembling a focal window of lens 1 that is much larger than that of lens 2. Moreover, the impinging proton trajectory is always horizontal. This is in fact the assumption we used to extract the trapping condition for displaced focal points in Eq. \eqref{eq:conditionDisplaced}. Then, replacing $w_{F1}$ by $\sigma_{\rm beam}\approx 50 \mum$ in Eq. \eqref{eq:Nwindow} we may find the trapping efficiency for the single-lens scheme. 
A major difference with respect to the double-lens scheme is precisely this dependency on the size of the beam spot.  As shown in Fig. \ref{fig:effL1}, the trapping efficiency is very sensitive to the ratio of focal windows $w_{F1}/w_{F2}$, now reinterpreted as $\sigma_{\rm beam}/w_{F2}$\footnote{In the double-lens scheme the proton beamlet is focused first and the dependency on $\sigma_{\rm beam}$ is indirect, entering through the focal distance $\Lone$, for a fixed $\Phi_1$.}.

\begin{figure}
	\centering
	\includegraphics[width=0.90\linewidth]{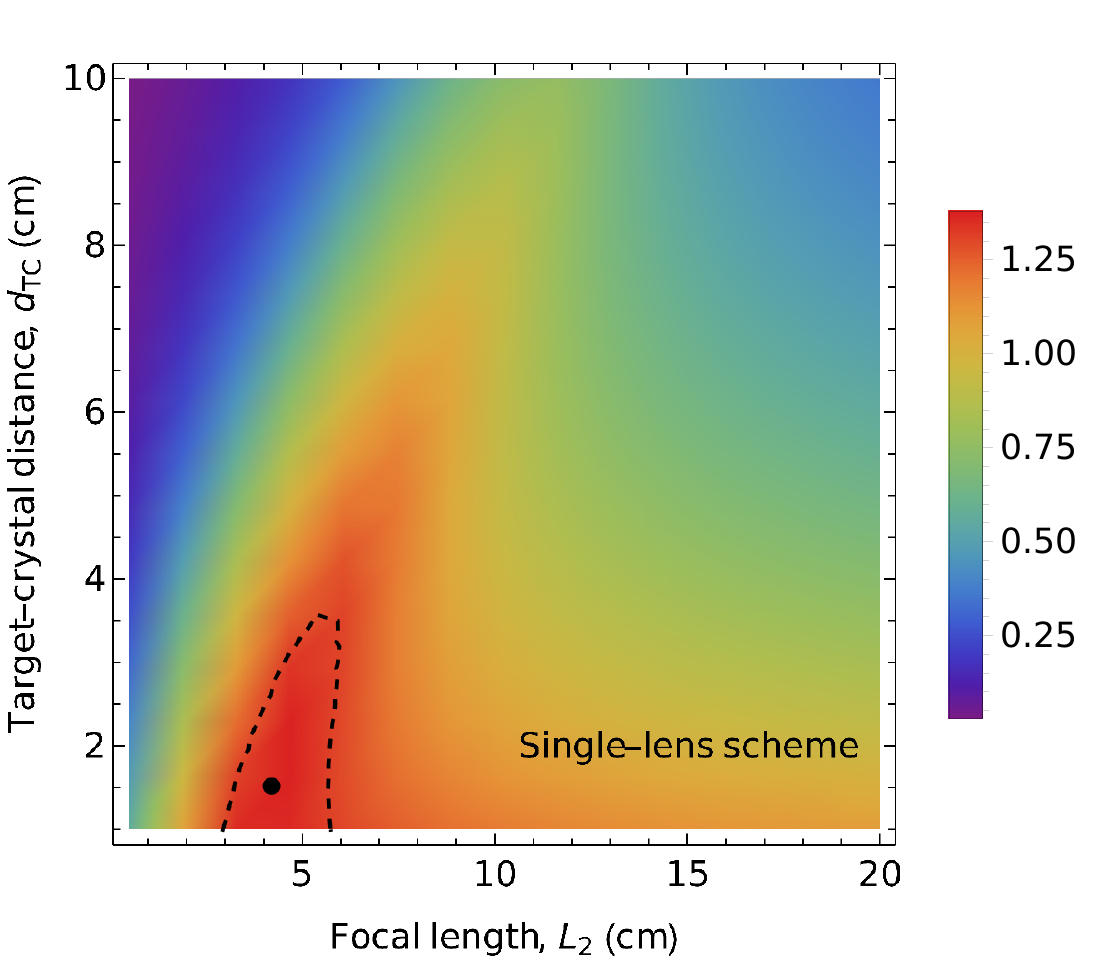}
	\caption{Efficiency gain of the single-lens scheme as a function of the target-crystal distance and the focal length. Besides the geometrical trapping condition, also the exponential decay of the $\Lc$ particles for variations of the the target-crystal distance is accounted for. The countour line encloses the region with a 5\% variation with respect to the optimal point, shown as a black dot.}
	\label{fig:singlelensoptim}
\end{figure}

The proton beamlet divergence is not affected in this scheme and there are no additional constraints on the setup parameters besides those already studied for the plain-crystal scheme. The only free parameters are the focal length $\Ltwo$ and the target-crystal distance $d_{TC}$, which previously were assumed identical. To decouple these parameters it is enough to change the integration region in Eq. \eqref{eq:ntrappedevents} from $r \in [-T/2, T/2]$ to $r \in [s -(T/2), s + (T/2)]$, where $T$ is the target thickness and $s$ is the separation from the m
iddle of the target to the focal point, $s=d_{TC}-L_2$. For a fixed focal length, the maximum of the trapping efficiency is found when the target is centered around the focal point, as expected from the enhanced efficiency at $r\to 0$ in Fig.~\ref{fig:phasespace} and the symmetry of the phase space around it. However, when fixing the the target-crystal distance first, the preferred position of the focal point is not in the middle of the target, but around 1 \cm before its center. This counter intuitive behavior can be explained by the increase on the focal window for larger focal lengths. Varying both parameters simultaneously, we find that the trapping efficiency gain from purely geometrical conditions increases monotonically with these parameters, reaching around a factor 7 for $L_2\approx d_{TC}\approx20\cm$. However, introducing the exponential decay of the $\Lc$ particles strongly reduces the trapping efficiency when the target-crystal distance departs from the reference value of 1\cm. The two-dimensional scan, shown in Fig. \ref{fig:singlelensoptim}, yields the optimal point at $\Ltwo = 4.2 \cm$ and $d_{TC}=1.5 \cm$, for which we find a 40\% increase in trapping efficiency with respect to the plain-crystal scheme.

In summary, while a substantial increase on the trapping efficiency is possible for larger target-crystal distances, this is overcompensated by the decay flight efficiency of the $\Lc$. For this reason, the single-lens scheme is better suited for dipole moment measurements of particles with longer decay times such as $\Xi_c^+$, $\Xi_b^+$ or $\Omega_b^+$.
Nevertheless, even if the improvement with this scheme for the $\Lc$ is modest, of 40\%, it needs no additional instrumentation with respect to the nominal layout besides the replacement of the plain crystal by a crystal lens, and it has no additional interference with the LHC beam.

\section{Technical challenges} \label{sec:technology}

The potential realization of the presented experimental configuration is subjected to the available crystal technology. The specifications for the needed crystal lenses, goniometers, and particle detectors are discussed in the following. 

\subsection{Crystal alignment} \label{sec:technologyAlignment}

A high precision goniometer is needed to align the first crystal lens to the beamlet of protons with a precision below the Lindhard angle, of $2 \murad$. This angular adjustment must be compensated in the vertical position of the second crystal lens, in order to make the focal windows overlap, ideally with a precision below 1 \mum, as shown in Fig. \ref{fig:trappingefficiency}. 
This accuracy is achievable with modern piezoelectric motors, which reach high positioning precision in the nanometer scale, retaining stability of position while unpowered. Such an inchworm technique was used \textit{e.g.} at RHIC to adjust the crystal in the beam \cite{rhic2005}. 
The required stability of the crystal position under beam exposure and related temperature shocks and radiation hits is similar to other crystal channeling applications at the LHC. In fact, the angular step of piezo-goniometers at the LHC is $0.1\murad$~\cite{lhc}, equivalent to a positioning step of $0.1\mum$ over a base of $1\,$m, which is even smaller than the precision needed for the double-lens scheme.
Thus, the losses due to the vertical displacement of the two crystals could in principle be avoided and the final efficiency gain is expected to be in the plateau region of Fig. \ref{fig:trappingefficiency}.

Regarding the angular alignment of the second lens, a precision of $100-200 \murad$ would be enough if the focal windows can be matched via the position of the crystal alone. An alternative may be to mount the setup with the target and two lenses offline and only adjust the angle of the whole structure, with a precision better than $2\murad$. 

In the single-lens scheme, the crystal alignment does not represent a critical point. The vertical position of the focal point should be within the beam spot, of $\sim 50 \mum$ in the transverse direction.

\subsection{Crystal dimensions and shape}

The first Silicon (Germanium) crystal prototypes following the specifications of the nominal layout, with plain faces, were built with a thickness of 5 (1) mm~\cite{Aiola:2020yam}. The bending was induced by clamping the crystals between the surfaces of a bender.
In our scheme, however, the thickness of the second crystal would only be $h_2 \approx L_2 \Phi_2 \approx 100\mum$, while the length remains the same, $\lengthtwo\approx 10\cm$.
In principle, it is possible to manufacture such thin crystals~\cite{Mazzolari:2014haa,PhysRevLett.108.014801} which can be bent through the deposition of a thin tensile or compressive film on one of its surfaces~\cite{GUIDI20111074}.

Alternatively, if working with a thicker crystal would ease the manufacturing or the experiment operation, a different shape for the crystal lens could be considered without affecting its performance: the circular cut to convert the plain crystal into a lens should only start at the middle of the entry face, covering the top part of the crystal entry face. In this way, only the top part of the crystal body would act as a lens, with fine faces of width $h_2$. The bottom of the crystal, with plain faces, could be arbitrarily thick, as it would not interact with the rest of the setup. The same arguments apply to the first lens. In that case, the minimum thickness of the focusing face is $h_1 = \sigma_{\rm beam} \approx 50 \mum$, and the cut should extend over the whole length of the crystal, achieving the triangular shape shown in Fig. \ref{fig:generalDiagram}. The length of the first lens $\lengthone$ does not have restrictions (Table \ref{tab:parameters}) and can be chosen to ease the manufacturing and reduce dechanneling loses.

\subsection{Reconstruction of spin precession}

The trapped particles in the second crystal enter within a range of initial directions of $\Phi_2 = 3.2\mrad$. This induces a spread on the deflection angle of the trapped particles in the range $[14,17] \mrad$. Consequently, the precession angle of the $\Lc$ magnetic moment, proportional to the deflection angle, changes depending on the channel followed by the particle. This spread, of $18\%$ would translate into a systematic error in the determination of the \Lc magnetic moment, probably of similar size, which is large compared to the statistical uncertainty of the planned experiment, of $\sim 4 \%$~\cite{Aiola:2020yam}. To reduce this uncertainty, it is necessary to be able to distinguish the path of the channeled $\Lc$. For instance, to reduce this uncertainty by a factor 5, the resolution on the $\Lc$ position at the crystal exit should be $ h_2 / 5\approx 20 \mum$. With the upgraded tracking system of the LHCb detector \cite{Collaboration:1624070,Collaboration:1647400}, this resolution is about $150 \mum $ \cite{Bagli:2017foe}, and some additional instrumentation after the crystal lens is required to achieve the needed resolution. In the scenario of a dedicated experiment to take place in a different location of the LHC ring~\cite{Bagli:2017foe,Aiola:2020yam,Mirarchi:2019vqi,Barschel:2020drr,Fomin:2019wuw}, the design of the tracking system should take this into account. 

A different solution to this problem can be found through the crystals rather than the particle detectors. The shape of the crystal entry face could be repeated at the crystal exit, but this time in a concave orientation, in such a way that all the particles undergo the same deflection angle (and thus the same precession). The spread of particle directions at the crystal exit would allow to precisely determine the channel of each particle due to the excellent angular resolution on these events \cite{Bagli:2017foe}. This would also allow to determine the initial $x$ component of the \Lc polarization for each event~\cite{Baryshevsky:2018hvh, Fomin:2019wuw}(defined as the direction perpendicular to the plane of magnetic precession, see \textit{e.g.} Ref. \cite{Aiola:2020yam}). The $x$ component, which is crucial for the sensitivity to the electric dipole moment, is very sensitive to the angle of the $\Lc$ with respect to the impinging proton. For this reason, in the plain-crystal scheme, a crystal tilt is introduced in the bending plane in order to collect $\Lc$ with some initial angle with respect to the horizontal. This is automatically realized by using crystal lenses, in which the particles enter with a spread of angles, leading to different $x$ components depending on the channel. In the double-lens scheme, even with the knowledge on the crystal channel of each particle, the initial angle cannot be determined accurately due to the unknown proton direction in each event. However, precisely because $\Phi_1$ has to be small compared to $\Phi_2$ (a limiting factor from the LHC beam pipe), the proton trajectory can be assumed to be horizontal to a very good approximation. In this solution, the data set could be split in bins of initial polarization along the $x$ axis, as allowed by the large increase in statistics, providing a method to keep the systematic uncertainties of the measurement under control.

\section{Conclusions} \label{sec:conclusion}

The experimental proposal to measure the electric and magnetic dipole moment of short-lived particles at the LHC suffers from very limited statistics. Two alternative layouts have been presented to improve the sensitivity of the measurement with the same number of protons on target. The trapping efficiency of the new experimental configurations has been evaluated in Section~\ref{sec:efficiency} by integrating over the phase space region that meets the trapping condition, carefully defined in Fig.~\ref{fig:channelingconditions}. The potential gain with the single-lens scheme can reach up to 40\% for $\Lc$ baryons, without any technical limitation besides the construction of the crystal lens itself, although this result is strongly dependent on the transverse size of the deflected beam. Instead, by introducing a short lens before the target to reduce this beam size, the statistical gain can reach up to a factor 20 in the double-lens scheme.
However, this solution distorts the shape of the proton beamlet and requires high accuracy on the relative position of the two crystals. These technical limitations were discussed in Section~\ref{sec:technology}.

Other factors affecting the sensitivity of the measurement such as dechanneling losses or the fit to a data set with different initial polarizations, are not expected to reduce the advantages of these experimental setups. Nevertheless, only Monte Carlo simulations reproducing the experiment with crystal lenses will set the final sensitivity gain of the single- and double-lens schemes. 

For the bottom baryons $\Xi_b^+$ and $\Omega_b^+$ a similar or larger statistical gain can be achieved due to their longer lifetime. 
Also, the improvement for the determination of the $\tau$-lepton magnetic moment is expected to be of similar size, but due to the additional complexity of its production, through $D_s^+ \to \tau^+ \nu_\tau $ decays, it should be evaluated with specific Monte Carlo simulations.


\acknowledgments

We are grateful to Fernando Martínez-Vidal and Nicola Neri for stimulating discussions and the revision of the manuscript. We also thank Adrea Mazzolari for valuable feedback.
JRV acknowledges support from MICINN and GVA (Spain), and the ERC Consolidator Grant SELDOM G.A. 771642.

\bibliographystyle{utphys}

\bibliography{biblio}

\end{document}